\documentclass[twocolumn,showpacs,amsmath,amssymb,pre,aps]{revtex4}   
\usepackage{graphicx}
\usepackage{dcolumn}
\usepackage{bm}
\usepackage{color}

\renewcommand{\vec}[1]{\mathbf{#1}}

\newif\ifgraph

\graphtrue             %

\begin{document}
\title{Activated barrier crossing dynamics of a Janus particle carrying cargo}

\author{Tanwi Debnath,\textit{$^{a}$} and Pulak Kumar Ghosh\textit{$^{b}$}\footnote[2]{Email: pulak.chem@presiuniv.ac.in}}

\affiliation{\textit{$^{a}$~Department of Chemistry, University of Calcutta, Kolkata 700009, India}}
\affiliation{\textit{$^{b}$~Department of Chemistry, Presidency University, Kolkata 700073, India}}

\date{\today}

\begin{abstract}
We numerically study the escape kinetics of a self-propelled Janus particle, carrying a cargo, from a meta-stable state. We assume that the cargo is attached to the Janus particle by a flexible harmonic spring. We take into account the effect of velocity field, created in the fluid due to movements of dimer's components, by considering space-dependent diffusion tensor (Oseen tensor). Our simulation results show that the synchronization between barrier crossing events and rotational relaxation process can enhance escape rate to a large extent. Also, the load carrying capability of a Janus particle is largely controlled by its rotational dynamics and self-propulsion velocity. Moreover, the hydrodynamic interaction, conspicuously, enhances the escape rate of the Janus-cargo dimer. The most of the important features in escape kinetics have been justified based on the analytic arguments.  
  
\end{abstract}
  \maketitle

\begin{figure}
\centering
\includegraphics[width=0.45\textwidth,height=0.20\textwidth]{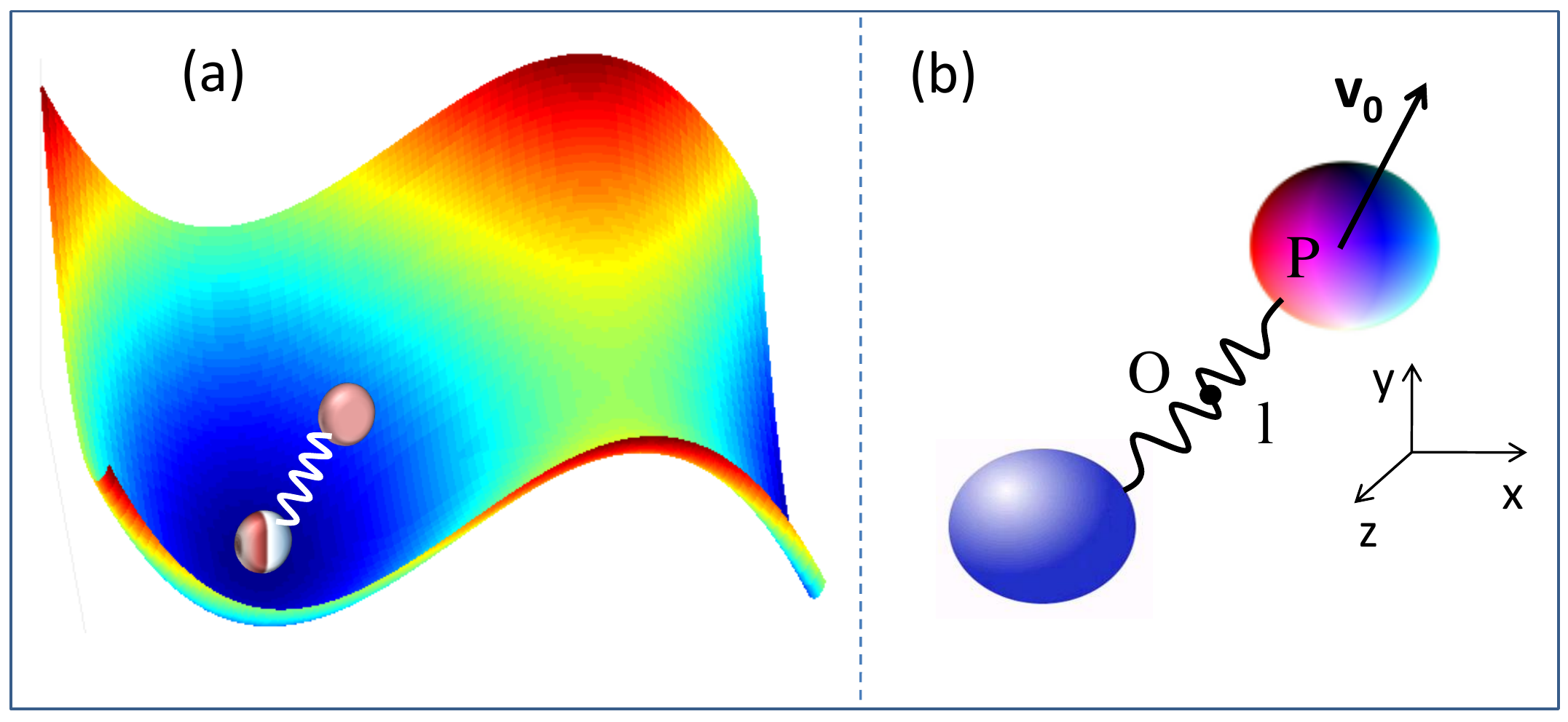}
  \caption{(Color online) (a) A schematic of meta-stable potential (2D cross-section of the 3D potential energy surface) with a Janus particle carrying a cargo at the bottom.  The meta-stable confinement has been modeled by the potential,
  $V(x,y,z)= \alpha x^2/2 - \beta x^3/3 + \lambda (y^2  + z^2)/2 $. (b) A schematic of active elastic dimer, O and P are respectively the center of mass and the center of force of the dimer formed by an active (Janus particle) and a cargo (passive particle) of equal radius, bound together by an elastic spring.  $\vec{{\rm v}_0}$ represents the instantaneous self-propulsion force\cite{forceVel} and $\vec{l} =l\hat{r}$ the dimers' length vector oriented from the passive to the active particle.}
  \label{fgr:example}
\end{figure}

\section{Introduction} Controlling motion of artificial micro-swimmers, like, self-propelled Janus particles (JPs), is a key issue to make best fitted Brownian tracer to use in medical sciences and nano-technology. Thinking of many other emerging and novel applications too, various types of self-propelled Janus particles have been synthesized\cite{review1,pccp0,cataly1,cataly2,cataly3}. Also, efforts have been devoted to characterize dynamical properties of self-propelled Janus particles which can be driven using light or chemical gradient. The main goal of these studies is to have a Brownian tracer  whose motion can be controlled upto a desired level of accuracy. Aiming at this, attentions have already been paid to understand self-propulsion mechanisms\cite{review2,review3, dsray}, behavior of JPs near the boundary \cite{fily1,volpe1,pccp2,our1},  collective behaviors \cite{review2, review3, Marchetti, Buttinoni},  autonomous motion \cite{our2, ai1}, interplay between shape and sizes during diffusion in confined geometries\cite{our3,ai4}, etc.

Microscopic details of self-propulsive mechanisms are quite complicated; however, it can be realized by considering effects of local gradients generated due to different physical or chemical processes taking place at the two distinct surfaces of Janus particles. For example, Janus particles having a hemisphere coated with catalyst can produce a local chemical gradient due to inhomogeneous reactions\cite{cataly1,cataly2,cataly3} . Also, inhomogeneous light absorption\cite{ther, pccp1} or magnetic excitation\cite{mag,sano2} of JPs can generate enough local temperature gradient for self-thermophoresis.

Dynamics of Janus particles, generally, is modeled by considering self-propulsion velocity vector parallel to the self-phoretic force. Further, the self-phoretic force keeps changing its direction due to fluctuations in local gradients or collisions with boundaries or the intrinsic rotational diffusion of the particle. Thus, the effective fluctuating force becomes time correlated and Janus particles exhibit active Brownian motion. Thanks to self-propulsion, Janus particles can diffuse orders of magnitude faster than a normal Brownian particles of the same size (e.g., a silica sphere of radius  2.13 mm,   half-coated with 20 nm thick gold caps, diffuses  $~ 7.84 \;\mu m^2$ per second  when the particle is exposed to a laser pulse of intensity 161 nW per micro meter square , where as a silica sphere of the same size (without gold coating) has diffusivity of  $~ 0.03\;\mu m^2$  per second  \cite{volpe1}). Transport properties of Janus particles are unusual as well as very interesting, e.g., they can exhibit autonomous motion when they encounter potential  (due to interaction with substrate) or confinement of asymmetric and periodic nature \cite{ai1, ai2, ai3, our2}, in some special non-equilibrium situations JPs can move opposite to the driving force \cite{GNM},  JPs can transiently drift towards the high fuel density or intense light  \cite{ourCHEMO, CHEMO1, CHEMO2, 
Photo}. All these features fascinate researchers to learn more precisely about motion of the particle so that they can be used in targeted drug delivery and other purposes in medical sciences.

To go forward to materialize the idea of using JPs as a drag carrier, in our previous study \cite{ourDimer} we explored some relevant issues in diffusion of Janus particles carrying a cargo in a shear flow. In this paper, we explore escape kinetics of a Janus particle carrying a load from an energetic trap modelled by a meta-stable potential. It is likely that a diffusing Janus particle on a surface of material encounters many local barriers that may appear in a periodic or random manner.  Moreover, the barrier height may be large or small with respect to the thermal energy. Therefore, in the context of drug delivery using JPs, it is desirable to have detailed knowledge about their escape kinetics from local minima. 

Our objective of the paper is twofold.  First, we intend to realize how escape rate is affected by the attached cargo, both, in the presence and the absence of hydrodynamic interaction (HI). Our second objective is to comprehend interplay between rotational dynamics, self-propulsion velocity and hydrodynamic interactions so that the parameter regime for better load carrying efficiency can easily be recognized. To this purpose, we numerically simulate Langevin dynamics of a Janus-cargo composite system to obtain escape rate as a function of key system parameters which could be tuned in experiments.  Prototype dimer systems comprised of two JPs \cite{exdimer1} or one JP and one passive particle\cite{exdimer2}, have already been synthesized to study their diffusion properties. Thus, our simulation results could be potentially important to guide experimentalist for suitably tailoring JPs as well as  attached cargoes and assigning desired self-propulsive properties.

The outlay of the paper is as follows.  In Sec. 2 we set up a model for the dimer Brownian dynamics that we implement in our numerical simulation code.  New results without and with hydrodynamic interaction have been reported in  Sec. 3.1 and 3.2, respectively.  Also, an overdamped rigid-dimer approximation of the model has been introduced in the beginning of the Sec. 3.1.  In Sec. 4, we conclude our study.

\section{ Model} We consider a self-propelled JP attached to a cargo (by chemical bonding) is diffusing in a three-dimensional (3D) cavity. The cavity is an energetic trap with potential energy (as shown in the Fig.1), 
\begin{eqnarray}\label{pot1}
V(x,y,z)= \alpha x^2/2 - \beta x^3/3 + \lambda (y^2  + z^2)/2.
\end{eqnarray}
The parameters $\alpha,\; \beta, \; {\rm and} \; \lambda  $  determine center, depth and width of the cavity. The dimer experiences a meta-stable potential, with a barrier,
\begin{eqnarray}\label{barrier1}
 \Delta V_0 \; = \; {\alpha}^3/6{\beta}^2,
\end{eqnarray}\label{pot1} 
while it moves along x-axis.  On the other hand, its movement along y and z directions is limited by harmonic potential well. Both, the JP and the cargo are of spherical shape with radius $a$. We model the chemical bonding between two components  in the Janus-Cargo dimer by a flexible harmonic spring with equilibrium bond length $l$. The interaction energy between the cargo and JP is, $ V_s(r_i,r_j) =  (k/2) (l-r_{ij})^2 $, where, $k$ is the measure of the coupling strength between two units of the dimer. 

In the overdamped limit, the displacement equations for two components of the dimer can be described as \cite{Ermak},
\begin{eqnarray}\label{1}
 r_i (t+\Delta t) = r_i(t) + \sum_{j}\frac{\partial D_{ij}}{\partial r_j}\Delta t + \sum_{j}\frac{D_{ij} {\rm v}_j}{D_0} \Delta t + \eta_i(\Delta t)
\end{eqnarray} 

where $r_i$ (with Cartesian components ($x_i, y_i, z_i$), and $i=1,2$) denotes the position of the i-th component's center of mass. The subscripts, $i=1 \;{\rm and  } \; 2$, correspond the active (Janus particle) and the cargo (passive particle) components, respectively. The drift velocity experienced by the Janus particles ($\vec{\rm{v}}_1$) and the cargo ($\vec{{\rm{v}}}_2$) is,
\begin{eqnarray}\label{2}
\vec{{\rm v}}_i =  - \vec{\nabla}_i V_s(r_i,r_j)/\gamma - \vec{\nabla}_i V(r_i)/\gamma + \delta_{i1}\vec{{\rm v}}_0.
\end{eqnarray} 
The first two terms of this equation arise due to harmonic interaction between two components of the dimer and substrate potential, respectively. The last term, $\vec{{\rm v}}_0$ is the self-propelled velocity which acts only on the active component of the dimer. As the self-propulsion directed to the certain symmetry axis of the Janus component, its direction with respect to the laboratory axis changes randomly due to the rotational diffusion.  Its  orientation fluctuation is governed by,
\begin{eqnarray}\label{psi}
{\bf {\dot {\hat p}}}= {\bf {\hat p}}\times {\bm \xi}(t),
\end{eqnarray}
but its modulus, v$_0$, is fixed. ${\bf {\hat p}}$ is the relevant unit vector and the three Cartesian components of the Gaussian noise ${\bm \xi}(t)$ are independent, zero-mean valued, and delta-correlated with variance $\langle \xi_i(\Delta t)^2\rangle = 2D_{r} \Delta t$. The raotational diffusion constant,  $D_r$  dictates the time decay of the ${\vec p}$ autocorrelation function \cite{cataly3}, $\langle p_q(t) p_{q'}(0) \rangle =(1/3)\delta_{qq'}\exp[{-2 D_r|t|}]$. It should be noted that rotational relaxation of self-propulsion is coupled to the dimer's rotations. In the rigid dimer limit, Eq.(\ref{psi}) is compatible with realistic situations when self-propulsion torque is zero (see Fig.2,  detailed discussions are in the sec. 3.1)


When mass and size of the segments of a dimer are larger than the surrounding solvent molecules, the host solvent mediated hydrodynamic interactions come into dynamics in a significant way. The moving JP ( or cargo) creates a velocity field which in turn affect the dynamics of the cargo (or JP). The effects of the hydrodynamic interaction enter into the dynamics through diffusion tensor in the equation (\ref{1}). The diffusion tensor is a nonlinear function of the instantaneous position of the all particles in the system. For a system of spherical particles it is possible to express diffusion tensor as a power series of inter-particle distances.  In lowest order approximation the diffusion tensor corresponds to an Oseen tensor, (and in the next to lowest order approximation is a Rotne-Prager tensor ). However, in the present problem, the diffusion of the dimer is led by the self-propulsive force acting upon the JP, while both monomers are subjected to a viscous drag. The self-propulsion mechanism (e.g., self-phoretic) may generate a short-range hydrodynamic backflow in the viscous suspension fluid \cite{Lauga}, which may result in an additional pair interaction between the monomers \cite{Popescu,ourJCPComm}. However, at low Reynolds numbers, as the active swimmer tows its cargo, it generates a laminar flow that tends to align the dimer's axis parallel to the propulsion force, with the PP trailing the JP. It is known \cite{Lauga,walls1,walls2} that the dipolar hydrodynamic interactions associated with the swimmer's propulsion mechanism decay faster with the distance than the perturbation due to the laminar flow caused by its steady translation. Therefore, the average tower-cargo distance can be conveniently chosen, so as to ignore the short-range hydrodynamic interaction between them with respect to the long-range hydrodynamic effects on their viscous drag.

We consider Oseen tensor, for the present problem, it is a (2 $\times $ 2) block component of (6 $\times $ 6) matrix of the following form, 
\begin{eqnarray}\label{tensor}
D_{ij} = D_0 I \delta_{ij}+ \frac{3D_0 a}{4r_{ij}}\left( I+\bar{r}_{ij}\bar{r}_{ij}/r^2_{ij}\right)
\end{eqnarray} 
The last term, $\eta_i(\Delta t)$, in the eq(\ref{1}), is a zero mean, Gaussially distributed random number with variance $\langle \eta_i(\Delta t) \eta_j(\Delta t) \rangle = 2 D_{ij}^0 \Delta t$. We refer\cite{Ermak} for detail about generation of the random number $\eta_i(\Delta t)$ in case of simulation.

The mechanisms of the translational and rotational diffusion may not be the same and therefore $D_0$, $\rm {v_0}$, and rotational relaxation time $\tau_r$ can be treated as independent model parameters. Finding analytic solution of the coupled  Eqs.(3-5) to obtain an expression for escape rate of the dimer from an energetic cavity is a formidable task. To accomplish our goal, we numerically solve\cite{Kloeden} Eqs.(3-5) to estimate the mean exit time ($T_{ {\rm MET}}$) of the active-passive dimer. The mean exit time is defined as an average time the dimer takes to exit the energetic trap starting from a random initial position around the bottom of the potential well. We also assume that the dimer axis and propulsive force may have  any orientation at the starting point of the simulation trajectories.   All the results (presented in the Figs.~3-5) are obtained by ensemble averaging over $10^4 -10^6$ trajectories depending upon the values of parameters. We choose times in seconds and lengths in microns.

\begin{figure}[h]
\centering
\includegraphics[width=0.35\textwidth,height=0.2\textwidth]{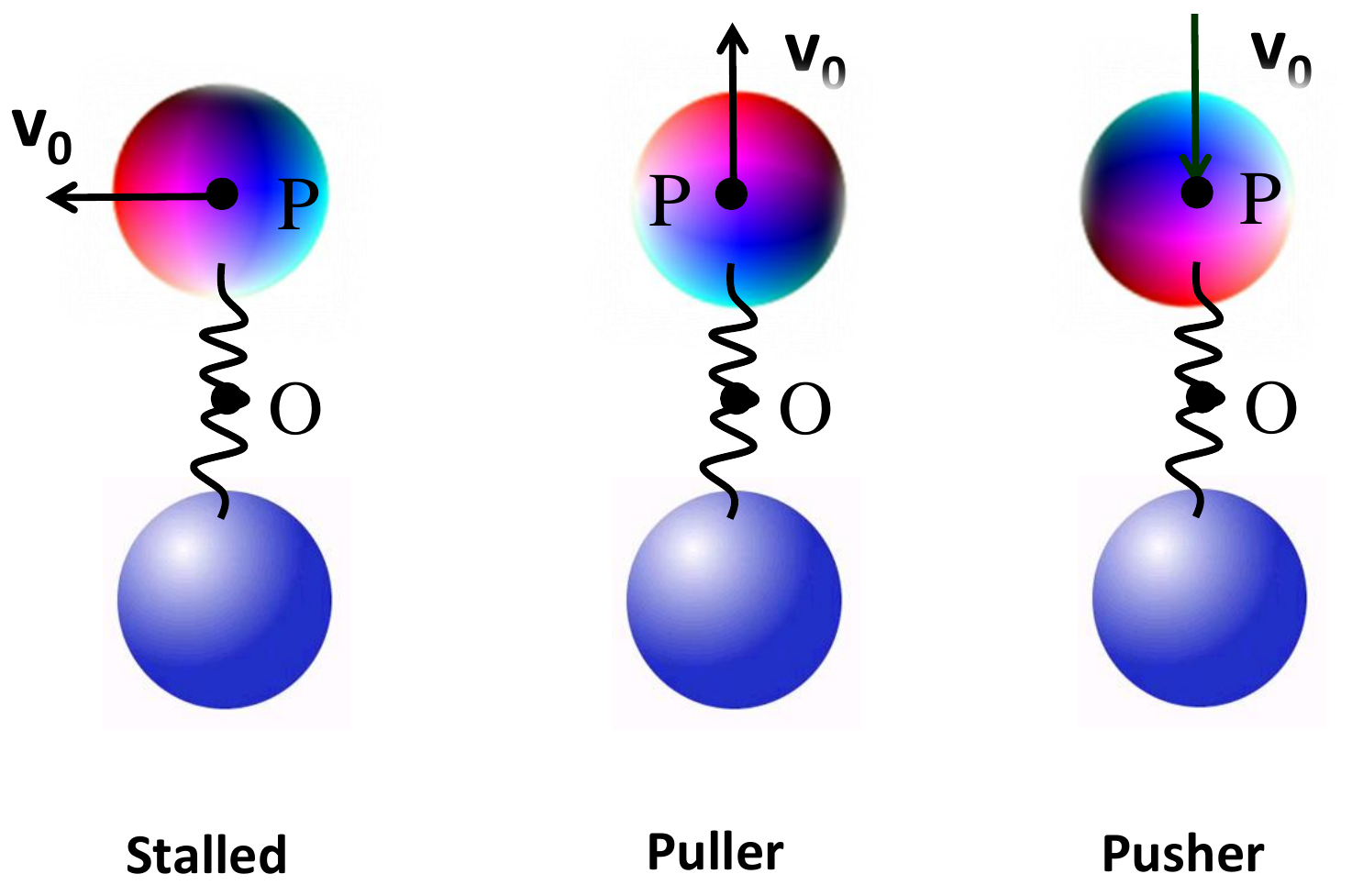}
\caption {(Color online) Schematics of three configurations of Janus-Cargo composite system.}
\end{figure}

\begin{figure}[h]
\centering
\includegraphics[width=0.40\textwidth,height=0.30\textwidth]{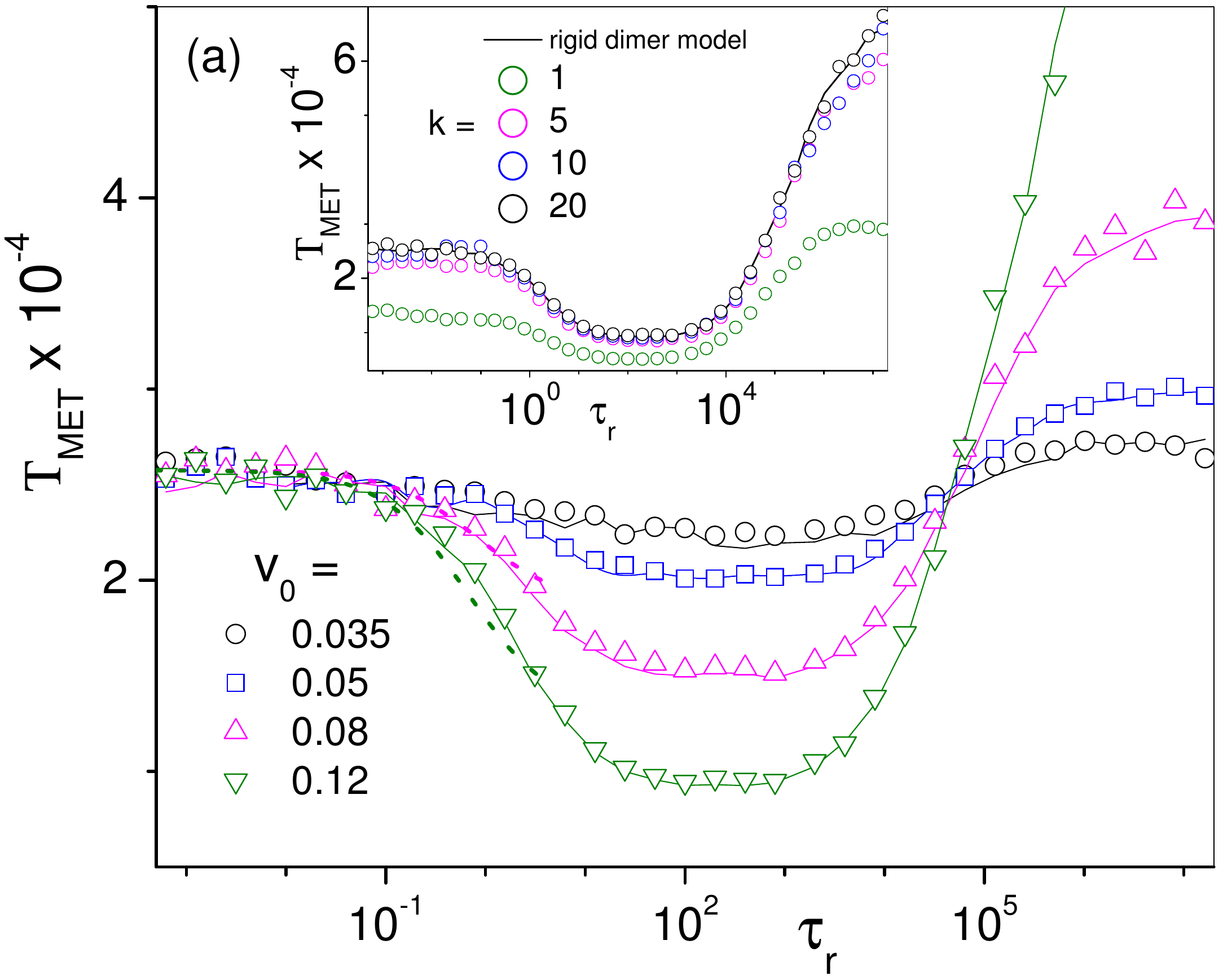}
\includegraphics[width=0.40\textwidth,height=0.30\textwidth]{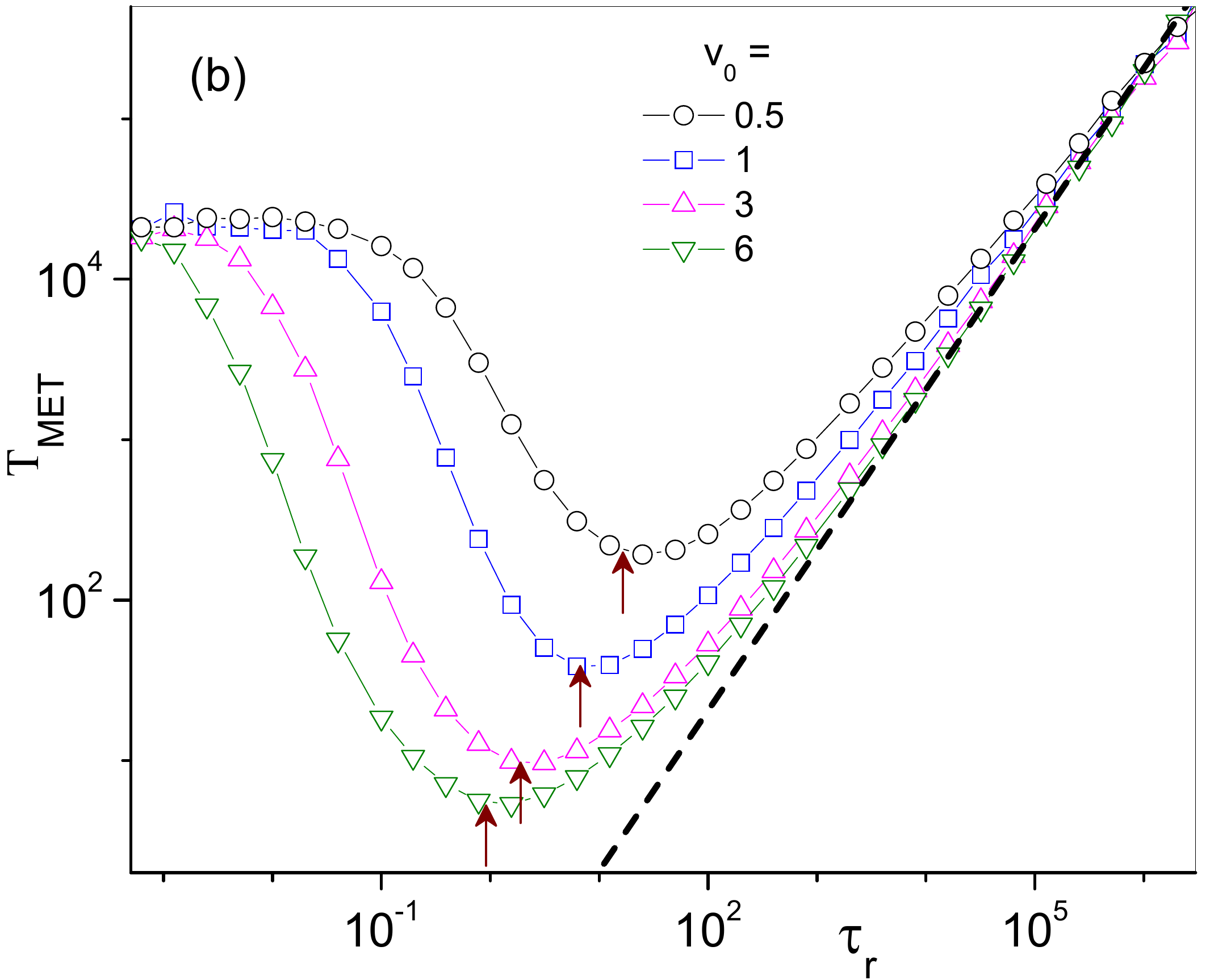}
\caption {(Color online) (a) Mean exit time ($T_{MET}$) as a function of rotational relaxation time for different values of self-propulsion velocity shown in the legends. Symbols and solid lines represents simulation of Eq.(\ref{1}) and Eq.(\ref{rigid-D}), respectively. Dotted lines for small $\tau_r$ are the estimation based on the Eq.(\ref{smalltau}). Parameter set used in simulation: $D_0 = 0.04, \; l = 0.1,\;  \alpha = \beta = 1,\; \lambda = 5,\; k = 20 \; {\rm and} \;a=0 \; {\rm (without \; hydrodynamic \; interaction)}$. {\it Inset: } $T_{MET}$ {\it vs} $\tau_r$ to examine validity of the rigid dimer approximation. Parameters are the same as main figure, except v$_0 = 0.12$  and the values of the spring constant are displayed in the legends. (b) $T_{MET}$ {\it vs} $\tau_r$ for large self-propulsion (comparable or larger than the barrier height $\Delta \tilde{V}$). Parameters used here are same as the Fig.3(a). Here the black dotted-line  represents linear growth of $T_{MET}$ with $\tau_r$ as per Eq.(\ref{liniarTau}) }
\end{figure}

\section{Escape kinetics of a Janus particle carrying a cargo}
As the escape kinetics of a Janus-cargo dimer from an energetic cavity is a challenging  one to study  analytically, we paid effort to explain our most of the simulation data in the framework of rigid dimer approximation. The rigid dimer limit is equivalent to the  non-Gaussian noise driven Brownian system as the self-propulsion force in case of Janus particles is distributed as,  $P({\rm v}_x) = {1}/{\pi \sqrt{{\rm v_0^2}-{\rm v^2}_x}}$, where v$_x$ is the x-component of v$_0$. Thinking of difficulties in handling the escape problem with non-Gaussian noise, we focus on the limiting situations where problem can be simplified considerably to have analytic estimates of the mean exit time.  The limiting cases arise depending upon the relative amplitudes of the three relevant time scales: (i) $\tau_k  $ -- time to retain the equilibrium configuration when a particle is slightly displaced from the bottom of the potential well, (ii) $T_{MET}^0$ --  exit time from the meta-stable state in the absence of the self-propulsion and (iii) $\tau_r $ -- the rotational relaxation time.  To this end, our study focuses on the  parameter regime where the restoring force  of the spring joining the components of Janus-cargo dimer is much stronger than the any relevant force in the dynamics. Based on the rigid dimer model, we analyze our simulation results, both, in the presence and absence of hydrodynamic interaction. In the Fig.~3--6 we present our key results which capture all important features of the escape kinetics. 
\begin{figure}
\centering
\includegraphics[width=0.40\textwidth,height=0.30\textwidth]{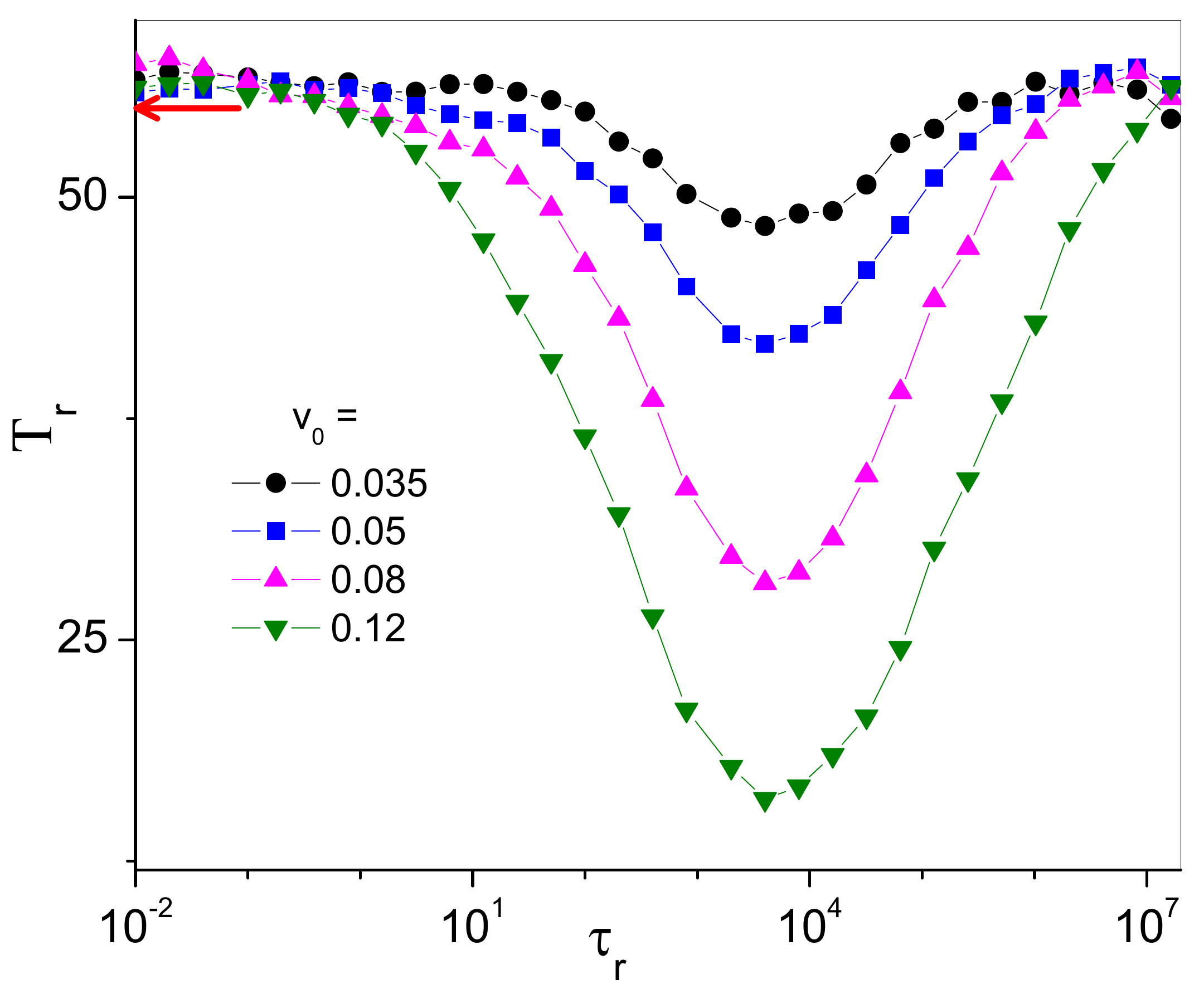}
\caption {(Color online)  $T_r (\equiv T_{MET}/T_S$) versus $\tau_r$ for different self-propulsion velocity shown in the legends.  Other parameters: $D_0 = 0.04,\; l = 0.1,\;  \alpha = \beta = 1 ,\; \lambda = 5, k = 20,\; {\rm and} \;a=0 \; {\rm (without \; hydrodynamic \; interaction)}$. The red horizontal arrow is a theoretical estimation based on Eq.(\ref{ratio1}) for $\tau  \rightarrow  0$.}
\end{figure}

\subsection{Mean exit time in the absence of hydrodynamic interactions}
Lets start with a relatively simple situation, the escape kinetics of Janus-cargo dimer in the absence of hydrodynamic interaction. To understand the interplay between time scales, we simulate $T_{MET}$ as function of rotational relaxation time with varying self-propulsion strength. As depicted in the Fig. 3(a), in the Kramers' regime the average escape time is independent of the rotational dynamics in the asymptotic limits, $\tau_r \rightarrow 0 \; {\rm and} \; \infty$. In the intermediate regime, all the plots passes through a flat minimum. The position of the minimum is quite insensitive to the self-propulsion velocity. On the other hand, beyond the Kramers regime, where self-propulsion can make barrier flat, different features emerge. For the large rotational relaxation time, the mean exit time grows linearly with $\tau_r$ and the position of the minima in  $T_{MET}$ versus $\tau_r$ linearly shift with self-propulsion strength [see Fig.3(b)]. To realize these interesting behaviors, we concentrate on the parameter regime where spring constant($k$)is strong enough to sustain thermal fluctuations, self-propulsion and other relevant forces without  appreciable change of the bond length. Thus, one can consider a fixed dimer length, $l$.

{\textbf {Rigid dimer model}} --- To set up the Langevin equation for the composite system under the rigid dimer approximation, we refer sketch in the Fig.1b. The center of force, P, and the center of mass ($c.m.$), O, are rest on the  dimer axis.
When the two monomers are identical, the modulus of the propulsion velocity $v_0$ of the JP is  related to the modulus of propulsion velocity on the composit system as, ${\rm v}_d ={\rm v}_0/2$. Moreover, under the action of self-propulsion force the dimer experiences the torque, ${\bm \tau}_p=(l/2) {\vec{\hat r}}\times {\vec {\rm {v_0}}}$, which tends to rotate the composite system around its center of mass. The thermal noises independently act to the two interacting components of the dimer. In the absence of hydrodynamic interaction due to the laminar flow around the two spheres, i.e., $a=0$, total six independent translational noise components $\eta_i(t)$  act on the rigid dimer through a translational noise \cite{Gardiner}, ${\bm \eta}_t(t)/\sqrt{2}$, and a fluctuating  torque ${\bm \tau}_\eta=(l/2) {\vec{\hat r}}\times {\bm \eta}_\tau(t)$. Here, the  components of ${\bm \eta}_t(t)$ and ${\bm \eta}_\tau(t)$ follow the same statistics as the $\eta_i(t)$ of Eq. (\ref{psi}). It should be noted that the random forces, $\eta_i (t)$, acting on a single monomer, cause a translational velocity $\eta_i(t)/2$ of the entire dimer; moreover, adding two independent $\eta_i (t)$ oriented along the same axis is statistically equivalent to replace them with another noise of twice variance  and identical statistics \cite{Gardiner}.

Based on the above analysis, the motion of center of mass, $O$, can be described by a 3D Langevin equation,
\begin{equation} \label{rigid-D}
\dot {\vec r}_O= - {\vec \nabla V_{cm}} + {\vec {\rm v}}_d +\sqrt{D_d}\;{\bm \xi}(t) ,
\end{equation}
where $r_O$ is the coordinate of the center of mass with Cartesian components $\{x_{O},\; y_O,\; z_O\}$ and orientation fluctuations of self-propelled velocity in the composite system, ${\vec {\rm v}}_d$, is entangled with rotational motion of the dimer.  In our model, the Janus particle's  c.m. coincides  with the center of the self-propulsion force. But, when a cargo is attached to the JP, the  center of  force lies apart  from the c.m.   (see Fig.1(b)). Thus, a self-propulsion torque ($\tau_p$) enters into the dynamics.  Following discussions in the previous paragraph, the orientation of dimer axis ($\hat r $) evolves in time as, 
\begin{equation} \label{rot1}
\dot {\hat r}= \frac{l}{2}\left(\hat r \times {\vec {\rm v_0}} +\hat r \times {\vec \eta_{\tau}} \right),
\end{equation}
$\vec {\rm v_0}$ itself can change due to rotational motion of the dimer around its own axis. The contribution of the first term (self-propulsion torque, $\tau_p $) in Eq.(\ref{rot1}) depends on the configuration of the composite system. For an analysis in detail about controlling factor of rotation motion, let's consider three configuration   shown in the Fig.~2. The stalled configuration, corresponds to the situation when self-propulsion acts perpendicular to the dimer axis, has maximum amplitude of self-propulsion torque ($\tau_p$). In general, the self-propulsion is much stronger than the thermal fluctuations, thus, $\tau_{\eta}$ can safely be ignored in this case. On the other hand, for puller and pusher configurations (angle between $\hat {r}$ and $\vec{{\rm v}_0}$ are, 0 and $\pi$, respectively) $\tau_p = 0$ and dimer rotation is solely determined by $\tau_{\eta}$. In such situation, $\vec{v}_d$ fluctuations can be described by the Eq. (\ref{psi}). It should be noted that, even in pusher or puller configurations, rotational diffusion of the JP-cargo system is different from the passive dimer system of the same size. This is likely to happen due to fluctuation of self-propelled velocity owing to spatial fluctuations of fuel density or light intensity, depending upon the self-propulsion mechanism. Thus,  relaxation time of $v_d$  can be considered as a free model parameter and related to dimer's rotation, 
\begin{equation}\label{rot2}
 \langle \hat r(t)\hat r(0) \rangle = \langle \vec{{\rm v}_d(t)} \vec{{\rm v}_d(0)}\rangle/\langle  {\rm v_d^2(0)}\rangle 
\end{equation} 
 Further, the Eq.(\ref{rot1})  assumes c.m. of the dimer rests at $l/2$. However, the c.m. and hence the amplitude of torques ($\tau_{\eta}\; {\rm and }\; \tau_{p}$)  may differ when cargo and JP are of different masses. For the sake of simplicity, our model considers cargo and JP are of the same size and mass. This offers us a simplified form of potential energy in terms of c.m. and relative distance between two components.  

The Cartesian components of the Gaussian noise ${\bm \xi}(t)$, in Eq.(\ref{rigid-D}), have mean $\langle \xi_i(t)\rangle =0$ and correlation functions $\langle \xi_i(t) \xi_j(0) \rangle = 2 \delta_{ij} \delta(t)$. As discussed above, for point like monomers with $a=0$, $D_d=D_0/2$.
The potential energy, $V_{cm}$, in the dynamics of the centre of mass is given by,
\begin{equation}\label{V-cm}
V_{cm}= \frac{1}{2}\left\{ \alpha x_O^2  + \lambda\left(y_O^2+z_O^2\right) \right\} - \beta \left(\frac{ x_O^3}{3}+\frac{ d_x^2 x_O}{4}\right), 
\end{equation}
It is similar to the substrate potential $V(x,y,z)$ experienced by the component of the dimer, but modified by an extra term, $d_x^2 x_O/4$. Where, $d_x = x_2-x_1$, is the x-component of separation between two monomers. It can safely be assumed one third of the equilibrium bond length $l$. Moreover, $d_x$ becomes equal to $l$ when potential energy surface is very stiff along y and z direction, i.e., $k >> \lambda >> \alpha $. In this situation, the dimer  preferably gets oriented along x-direction with a fixed bond length rest at its equilibrium value.  

Based on the rigid dimer model, the diffusion properties of the active-passive composite system are detailed in Ref. \cite{DDsoft, ourDimer}. For the sake of simplicity, if suffices to recall here that the noise strength $D_r$ plays the role of an orientational diffusion constant, whose inverse, $\tau_r$, quantifies the temporal persistency of the {\it isotropic} Brownian motion of the dimer's center of mass. To check validity of the rigid dimer approximation, we compare simulation results produced from Eq. (\ref{rigid-D}) with Eq.(\ref{1}). It is apparent from Fig.3a (inset), as soon as the spring constant grows stronger, the rigid dimer model appears as a good approximation. In Fig.3a and 3b, we depict variation of mean exit time as a function of $\tau_r$ for the Kramer's regime and beyond it, respectively. To analyze simulation results, we separately focus on the three regimes of the rotational relaxation time.

\noindent{\textbf{ Fast rotational relaxation }}  --- When the rotational relaxation time is much faster than the other relevant time scales,  $T_{MET}^0$ and  $ \tau_k$,  the effect of self-propulsion amounts to enhancement of diffusivity with a modified temperature $T_{eff}$. Such situation is equivalent to  passive particle's  activated barrier crossing dynamics with an effective temperature \cite{temp1, fixedangle},  
\begin{eqnarray}\label{teff2}
T_{eff} =  T\left[1+{v_d^2 \tau_r}/{2D_d\left(\tilde{\alpha} \tau_r+1\right)}\right]. 
\end{eqnarray} 
Where, $\tilde{\alpha} = \alpha/\gamma$, determines the amplitude of drift velocity a particle experiences when slightly lifted from the bottom of the meta-stable potential. Thus, the mean exit time can be expressed as \cite{Kramers2, Kramers1},
\begin{eqnarray}\label{smalltau}
T_{MET} =  \frac{2\pi}{\omega_t \omega_b }\exp{\left[\frac{\Delta \tilde{V}}{k_B T_{eff}}\right]}
\end{eqnarray}  
where, $\omega_t$ and $\omega_b $ are the frequencies at the barrier top and  bottom of the meta-stable potential (\ref{V-cm}), respectively.  The barrier height of the same potential can be approximated as, $\Delta \tilde{V}  \approx\Delta V_0 - \beta d_x^2x_m/4$, where $x_m$ is the position of barrier top. The Eq. (\ref{smalltau}) corroborates our simulation results. For a sake of comparison, our estimates from Eq.(\ref{smalltau}) have been displayed by dotted lines in Fig.3a. It is apparent from, both, the figures  and the equation, when $\tau_r$ tends to zero, $T_{MET}$ becomes insensitive to the rotational dynamics and approaches to $T_{MET}^0$. On increasing relaxation time, the effect of self-propulsion appears noticeable as soon as, $\tau_r\geq 2D_d/(v_d^2 - 2D_d\tilde{\alpha})$.  Due to enhancement of effective temperature, the mean exit time first decreases with $\tau_r$, then grows almost linearly passing through a minimum. This behavior cannot be explained based on Eq(\ref{smalltau}) as it looses its validity as soon as the rotational time scale approaches to  $\tau_k $ or $ T_{MET}^0$.

{\textbf{ Very slow rotational relaxation}} --- We now consider the situation when the rotational relaxation time is much larger than any relevant time scale in the system. In this regime two apparently distinguishable behaviors are observed. Firstly, up to a certain self-propulsion strength, the mean exit time reaches to a saturation limit at large $\tau_r$. The asymptotic value depends on the $v_0$.  Secondly, for large $v_0$, when the perturbation amplitude due to self-propulsion is comparable  to the barrier height of the unperturbed system, the mean exit time linearly grows with $\tau_r$. To understand this feature we resort to the fixed angle approximation\cite{fixedangle}. It assumes, in the large   $\tau
_r$ limit, the escaping particle experiences a constant potential,
\begin{eqnarray}\label{angle1}
U(x_O,\theta) =  V_{cm}(x_O) - v_d \gamma x_O\cos{\theta},
\end{eqnarray}
with a parametric dependence on the direction of self-propulsion, $\theta$. Based on the conventional flux over population method,  $T_{MET}$ can be expressed as a function of $\theta$, 
\begin{eqnarray}\label{angle2}
\tilde{T}(\theta) =  \frac{2\pi}{\sqrt{|U''(x_{t'})||U''(x_{b'})|}}\exp{\left[\frac{U(x_{t'})-U(x_{b'})}{k_b T}\right]}
\end{eqnarray}
Where, $x_{t'}$ and $x_{b'}$ are the top and bottom of the effective potential, $U(x_o,\theta)$.  Note that the Kramer's like rate equation(\ref{angle2}) is valid only when the perturbation in potential due to self-propulsion much smaller than the barrier height of the unperturbed  system, i.e., $v_0 \gamma(x_{t'} -x_{b'}) \ll \Delta V_0$. The final rate expression can be obtained  by taking average over $\theta$ (with an uniform distribution over the range $0$ to $2\pi$). 
\begin{eqnarray}\label{angle3}
 T_{MET} =  \langle \tilde{T}(\theta)\rangle
\end{eqnarray}  
To help interpret simulation results, we define two more exit times corresponding to the tilting of the potential to the left and right. The self-propulsion may assist (hinder) the exit process by  setting lower, $V_{-}$, (higher, $V_{+}$,) potential barrier. Let us consider, $T_{+}$ and $T_{-}$ are  the mean exit time of the trajectories who encounter higher and lower  barriers, respectively. Quantitatively, they can be defined as average value of $\tilde{T}(\theta)$ over the range $-\pi/2 $ to $ \pi/2$ and $\pi/2 $ to $ 3\pi/2$, respectively. 
It is evident from Eq.(\ref{angle1} - \ref{angle2}) that $T_{+}$ is an increasing,  while  $T_{-}$ is a decaying function of $\rm {v_0}$. As soon as $T_{+}$ approaches to $\tau_r$, the path with higher barrier starts becoming inefficient in barrier crossing process. Thus, $T_{MET}$ increases with $\rm {v_0}$. When, $T_{+}$ grows quite larger than $\tau_r$, only in the interlude having lower barrier, the particle can escape the trap. Thus, $\rm {v_0}$ dependence in $T_{MET}$ appears through $T_{-}$.  Hence, $T_{MET}$ suppresses with modulus of propulsion velocity and turn out to be insensitive   of $\rm {v_0}$ for $T_{-} \ll \tau_r$. 

On the other hand, for quite a large $\rm {v_0}$ (beyond the Kramers' regime, see Fig.3(b)) and $T_{+} > \tau_r$, the average exit time grows linearly with $\tau_r$. This is due to the fact that in a large fraction of the trajectories, the Janus swimmers need to wait for acquiring propulsion in the appropriate direction (by rotation motion) to move towards the basin. On an average, 50\% trajectories start with initial x-component velocity negative and the rest have positive direction. If propulsion is large, the half of the trajectories with initial propulsion direction $\theta = \pi/2 \;{\rm to} \; 3\pi/2$ exit in two steps: first, the particles diffuse to acquire positive x-component of propulsion velocity. Required time for this is, $\pi^2/12D_r$. In the second step, the particles move to the absorbing point with average velocity $2\rm {v_d}/\pi$. Thus, the rate determining 50\% trajectories have average exit time, $\pi^2/12D_r + \pi|x_m -x_a|/2\rm {v_d}$. Taking into account exit time of another 50\% trajectories, 
\begin{eqnarray}\label{liniarTau}
T_{MET}={\pi^2 \tau_r}/{24} + {\pi |x_m -x_a|}/{2\rm {v_d}}
\end{eqnarray}
For large $\rm {v_d}$ and $\tau_r\rightarrow \infty $, the second term of the above equation is insignificant.  Simulation results in  Fig.~2b corroborate well with the above estimation (shown by dashed lines).   

\noindent{\textbf{ Intermediate regime of rotational relaxation time}} ---  The most interesting  features are observed in escape kinetics when the rotation relaxation time is of the order of $ T_{MET}^0$. This regime cannot  be realized by neither adiabatic approximation nor white noise approximation.  Recall that for very fast rotational relaxation, the fluctuations of effective barrier height due to self-propulsion are average out in the time scale of the order of $T_{MET}^0$.  In the opposite limit, $\tau_r\gg T_{MET}^0$, particles need to wait until the direction of propulsion force gets reversed by rotational diffusion. In the intermediate regime, a synchronization between thermal noise-induced (or propulsion driven) hopping and rotational relaxation is manifested by a resonance like behavior in $T_{MET}$ versus $\tau_r$ (see Fig.3 ). This type of resonance phenomena is known as { \it resonant activation}\cite{resonant} in literature.
 The position of the minima in the Fig.3a and Fig.3b, can be understood with some simple arguments. First, we consider the limit when perturbation due to self-propulsion is larger than $\Delta V_0$, {i.e.} beyond the Kramer's regime. In such situation, the particles can cross the barrier only when self-propulsion is directed in the range, $\theta = -\pi/2 \;{\rm to} \; +\pi/2$. As soon as the particle acquires suitable direction, it drifts toward the absorbing point ($x_a$) with an average velocity, $2\rm {v_d}/\pi - \bar{v}_d$. Thus, a perfect synchronization between rotational dynamics and escape process requires, 
\begin{eqnarray}\label{resonance}
(\tau_r)_m = \frac{\pi|x_m - x_a|}{2\rm {v_d} - \pi \bar{v}_d}
\end{eqnarray}  
Where, $\bar{v}_d$ is the thermal average drift speed in the potential valley in between top and bottom of the unperturbed system. For a harmonic potential well, $\bar{v}_d = \langle \alpha x/\gamma\rangle \approx \sqrt{2k_B T \alpha/\pi \gamma^2}$. Based on the equation(\ref{resonance}) positions of the minima can be predicted for $\rm {v_0} \gg \bar{v}_d$. The predicted values are in accord with simulation data (indicated by vertical arrows in Fig.3b). In the weak self-propulsion limit, $\rm {v_0} \gamma (x_t -x_b) \ll \Delta V_0$ the depth of the minima in $\tau_r$ vs $T_{MET}$ diminishes and the resonance becomes feeble (even disappears). It appears from the simulation results that the synchronization between rotational dynamics and fluctuation-induced barrier crossing event is perfect when  $\tau_r$ approaches to lowest estimated escape rate as per the equation (\ref{angle2}). It corresponds to one of the most probable value of the self-propulsion.

\noindent{\textbf{Effects of load in the escape kinetics}} ---
We conclude this section with an analysis of load carrying capability of Janus particles as a function of rotational relaxation time. To this purpose we simulate the ratio, $ T_r = T_{MET}/T_{S}$ as a function of $\tau_r$ with varying self-propelled velocity. Where, $T_S$ is the mean escape time of a single JP without cargo.  The inverse of $T_r$ can be treated as a quantifier of load carrying capability of the JP. Figure 3 shows that both the asymptotic limits reach a constant value which is independent of v$_0$. There is an intermediate regime of $\tau_r$ where the ratio attends its reaches. Noticeably, the position of the minimum is independent of the propulsion velocity. We recall the Eq.(\ref{teff2}) to explain the features of Fig.4 at very short $\tau_r$. Simplifications of the estimate based on the assumption of effective temperature produces,     
\begin{eqnarray}\label{ratio1}
T_{r} =  \exp{\left[ \frac{8\Delta \tilde{V}}{4D_0+{\rm v}_0^2\tau_r}- \frac{2\Delta V_0}{2D_0+{\rm v}_0^2\tau_r}\right]}. 
\end{eqnarray} 
 In the limit, $\tau_r \rightarrow \; 0 $, the Eq.(\ref{ratio1}) reduces to,
\begin{eqnarray}\label{ratio2}
T_{r} \sim  \exp{\left[ ({4\Delta V_0-\beta x_m d_x^2})/{4D_0}\right]}. 
\end{eqnarray} 
This equation is in accord with simulation results (indicated by a horizontal arrow in Fig.4).

 A similar asymptotic behavior is observed in the opposite limit, $\tau_r \rightarrow \infty$, $T_{r}$ remains unchanged with variation of $\tau_r$.  Moreover, noticeably, the asymptotic values of load carrying capability are independent of v$_0$. Beyond the Kramers' regime, such type of feature  can be understood based on the asymptotic behaviors of mean exit time of the both, dimer, as well as, the single JP. As it is evident from previous analysis, for the both cases $T_{MET}$ linearly grows with $\tau_r$. Thus, the nature of variation of  $T_r$, at $\tau_r \rightarrow \infty $ is quite explicable.

Most conspicuously, the Janus-cargo dimer can cross the barrier with its fastest rate at some $\tau_r$ value  which fall quite a far away from the resonance positions either of the single JP or Janus-cargo composite system. The position of the minima in $T_r$ versus $\tau_r$ is independent of the v$_0$. However, the depth of the minimum (see Fig.4) has a very strong monotonic dependence of the self-propulsion velocity. For the parameter set used in the Fig.4, the escape rate can be upto four times faster than a normal Brownian particle. Thus, our study indicates that Janus particle have a potential to carry load with quite a large efficiency.

\subsection{Effects of hydrodynamic interaction in the escape kinetics}
To examine the effects of hydrodynamic interaction in the barrier crossing dynamics we simulate mean escape time as a function of self-propulsion strength and size of the monomers for different rotational relaxation times. They have been depicted in the Fig.5 and Fig.6. Careful inspection of the simulation results (in Fig.5) reveals  that all the distinct features of rigid dimer's escape kinetics are retained in the presence of the hydrodynamic interaction. Particularly, like single Janus particles (equivalent to the rigid dimer), interplay among  time scales, $\tau_r$ and $T_{MET}^0$, apparently manifested in the asymptotic behaviors of escape time and its parametric dependence. Further, it is evident from the simulation data that hydrodynamic interaction considerably facilitates the escape rate of Janus-cargo dimer. We quantify the effect of hydrodynamic by estimating the ratio,  $T_h = T_{MET} /T_{MET} (0)$. $T_{MET} (0)$ denotes here the numerically simulated mean escape time for the same parameter as $T_{MET}$ but  $a=0$. The escape kinetics is free from hydrodynamic interaction when $T_h = 1$, and $T_h > 1$ and $T_h < 1$, correspond to the negative and positive impacts, respectively.  The dependence of the ratio on v$_0$ for different particle sizes and rotational diffusion times have been displayed in Fig.5(a, b). For small v$_0$, as long as the translational diffusion is larger than the self-propulsion contribution, both the mean exit time and the ratio $T_h$ are insensitive to the self-propulsion velocity. Simulation results show that the onset of the noticeable self-propulsion effects is determined by rotational relaxation time and particle size for a fixed $D_0$. The effect of the hydrodynamic interaction is prominent when v$_0$ is low and disappear when propulsion is large (see Fig.5(b)). We try to understand all these interesting features based on the following analysis.

The hydrodynamic effects have been incorporated by taking into account the spatial dependence of the diffusion tensor in Eq. (\ref{tensor}). Considering leading order \cite{Ermak}, the modification of the thermal noise can be written as, 
\begin{figure}
\centering
\includegraphics[width=0.45\textwidth,height=0.40\textwidth]{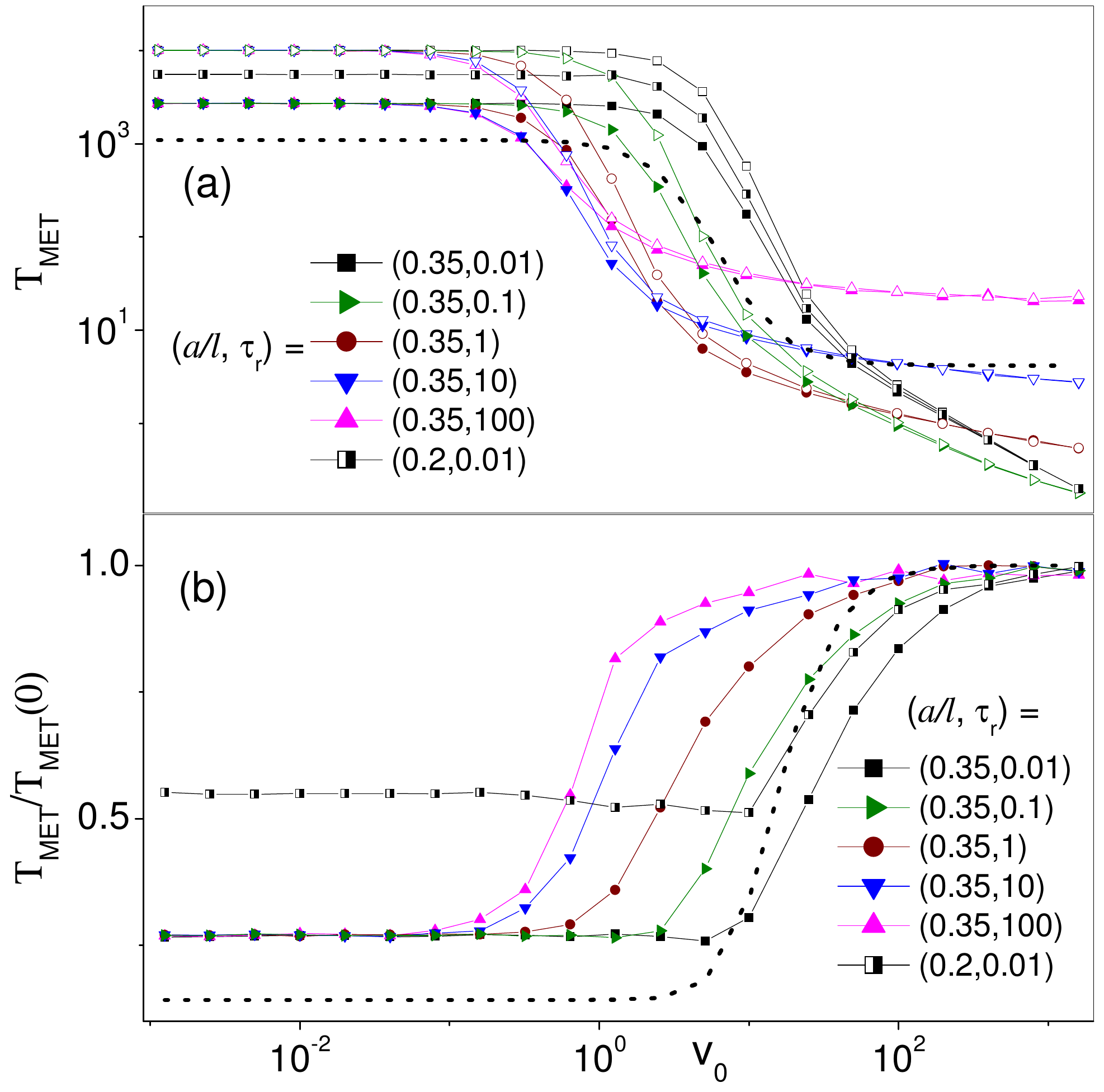}
\caption {(Color online) (a) Mean exit time ($T_{MET}$) of active-cargo dimer from the meta-stable state as a function of self-propelled velocity for different rotational relaxation times and  particle sizes (mentioned in the legends). Lines with solid symbols display simulation results with hydrodynamic interaction, where as the hollow symbols depict the same but in the absence of hydrodynamic interaction.  Simulation parameters are $D_0 = 0.14,\; l = 0.3,\;  \alpha = 1.5,\; \beta = 1 ,\; \lambda = 5, {\rm and} \; k = 20 $. Dotted line is the prediction from our analytical result Eq.(\ref{teff4}) for $\tau_r \rightarrow 0$.  (b) The ratio $T_h  ( \equiv T_{MET} /T_{MET} (0))$ versus v$_0$ for different $\tau_r$ and $a$. Simulation parameters are the same as the subfigure (a). The dotted line is the estimation based on the Eq.(\ref{teff4}).}
\end{figure}
\begin{equation}\label{D0a}
D_0 \to D_0(1+a/l).
\end{equation}
Further, in the Eq. (\ref{1}), the force term too has been multiplied by $D_{ij}$, which implies that, in the same order of $a/l$, the self-propulsion velocity must also be rescaled as
\begin{equation}\label{v0a}
{\rm v_0} \to {\rm v_0}(1+a/l).
\end{equation}
\begin{figure}
\centering
\includegraphics[width=0.45\textwidth,height=0.30\textwidth]{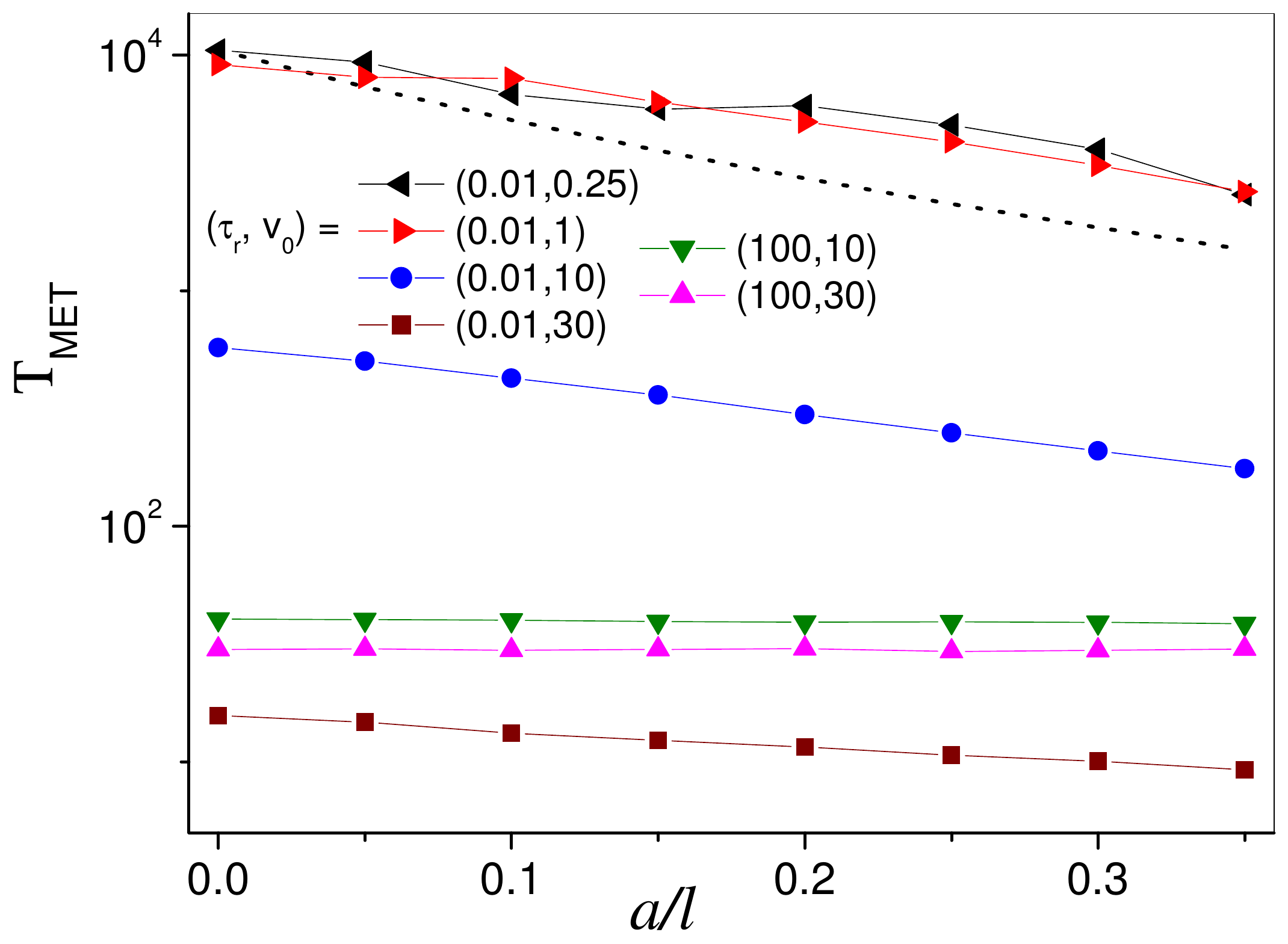}
\caption {(Color online)  $T_{MET}$ as a function of $a$  for different rotational relaxation times and  self-propelled velocities as mentioned in the legends. Simulation parameters are $D_0 = 0.14,\; l = 0.3,\;  \alpha = 1.5,\; \beta = 1 ,\; \lambda = 5,\; {\rm and} \; k = 20 $. Dotted line is the prediction based on the Eq.(\ref{teff4}) for $\tau_r \rightarrow 0$.}
\end{figure}
The hydrodynamic corrections to the translational diffusion and self-propulsion lead to modification of diffusivity of the composite system and hence effective temperature which activate the dimer to cross the barrier. The effective temperature can be expressed as,     
\begin{eqnarray} \label{teff4}
T_{eff} =  \frac{D_d(1+a/l)}{k_B}\left[1+\frac{v_d^2(1+a/l)^2 \tau_r}{2D_d(1+a/l)\left(\tilde{\alpha} \tau_r+1\right)}\right]. 
\end{eqnarray}
As mentioned earlier,  in the context of escape kinetics, the concept of effective temperature is meaningful when the $\tau_r$ is much smaller than any relevant time scale. Thus, using Eq.(\ref{teff2}) and Eq(\ref{smalltau}) one can capture some features of $T_{h}$ versus v$_d$ shown in the Fig.5(b). In case of passive dimer, v$_d=0$, 
\begin{eqnarray} \label{teff5}
T_{h} =  \exp{\left[-\frac{\Delta \tilde{V}}{D_d\left(1+l/a \right)} \right]}. 
\end{eqnarray}
This value can be compared with the simulation results. In the opposite limit, $D_0 = 0$ or $D_0 \ll {\rm v}_d$,
\begin{eqnarray} \label{teff6}
T_{h} =  \exp{\left[\frac{2\Delta \tilde{V}}{{\rm v}_d^2 \tau_r} \left(\frac{1}{(1+a/l)^2} - 1\right)\right]}. 
\end{eqnarray}

\begin{table*}
\small
  \caption{Summary of parametric dependence of $T_{MET}$ in the various limiting situations of rigid dimer}
  \label{tbl:example}
  \begin{tabular*}{\textwidth}{@{\extracolsep{\fill}}|c|c|c|c|}
    \hline
    --- & Fast rotational relaxation, & Intermediate regime of & Slow rotational relaxation, \\
        & $T_{MET}^0\gg \tau_r$       & rotational relaxation, & $T_{MET}^0\ll \tau_r$ \\
        &                             & $T_{MET}^0\sim \tau_r$  &                       \\
    \hline
    $T_{MET}$ without HI in & Reaches to the limiting value, & Position of minima in $T_{MET}^0$ vs & Reaches to a v$_0$ dependent \\
    the Kramers' regime . & $T_{MET}^0$, which is independent of & $\tau_r$  is independent of v$_0$ but its & asymptotic value. \\
                & both v$_0$ and $\tau_r$. & dept is a monotonic function of v$_0$. &                   \\ 
    \hline
    $T_{MET}$ without HI beyond & Like in Kramers' regime, $T_{MET}$ & Position of v$_0$ dependent & Grows linearly with $\tau_r$ as \\
    the Kramers' regime. & reaches to a v$_0$ independent & minima can be predicted & per Eq. (16). \\
                & asymptotic value. & using Eq. (17). &                 \\  
    \hline
     Effect of HI on exit & Features are similar to without HI, & Position of minimum remains & In Kramers' regime, $T_{MET}$ \\
     time. & but $T_{MET}$ is reduced to a large extent. & unaltered   both in the & reaches to a asymptotic       
     \\
                & However, for large v$_0$ effect of HI & Kramers' regime and beyond & value which is function of \\
                & disappears. & it. & both, v$_0$ and $a$. \\
      \hline

    
  \end{tabular*}
\\ Note: $T_{MET}$ -- mean exit time, $T_{MET}^0$ -- mean exit time in absence of self-propulsion, HI -- hydrodynamic interaction, v$_0$ -- self-propelled velocity, $a$ --particle size. \\
\end{table*}

In the low self-propulsion limit dynamics is guided by the thermal fluctuations and $T_h$ becomes insensitive to the both v$_0$ and $\tau_r$. As soon as the self-propulsion contribution surpasses the translational diffusion, $T_h$ starts increasing very slowly with increase of v$_0$ or $\tau_r$. After a certain value of v$_d$, the effect of hydrodynamic interaction starts getting weaker and almost disappear at v$_d \rightarrow \;  \infty$. This type of behavior can be understandable recalling that the dimer  length $l$ is not fixed. The force acting on the one monomer elongate equilibrium bond length. Neglecting rotational corrections and assuming steady propulsion conditions, the force balance condition yields, $l(\rm {v_0}) = l + \rm {v_0}/2k$. As a result, on increasing $\rm {v_0}$ the ratio $a/l(\rm {v_0})$ tends to vanish and so does the hydrodynamic effect disappear on the active dimer's escape kinetics.

Figure 6 depicts a very slow enhancement of escape rate with increase of the particle size in the fast rotational relaxation regime. However, for large self-propulsion and slow rotational dynamics the escape kinetics becomes insensitive to the hydrodynamics. Our analytic estimates qualitatively corroborate all these features of simulation results (see Fig.5(b) and Fig.6). However, discrepancy in quantitative level may be attributed to the slight modification of potential energy surface. This is expected as the meta-stable potential contributes to the force term in the Eq. (\ref{1}) which has been multiplied by $D_{ij}$. 

  Figure 5 and 6 give a clear picture on interplay between hydrodynamic interaction and rotational relaxation time. In the rigid dimer limit, for a given self-propulsion strength, $T_h$ is a monotonically deceasing function of $\tau_r$. This may imply that for fast rotational motion, components of dimer are quick enough to be arranged in the velocity field in such a way that enhances diffusion of the composite system in comparison to the same in the absence of the HI. Further, for very large self-propulsion limit the movement of passive particles guided by directional motion (persists up to a distance  $\tau_r v_0 / 2$), but the equilibrium distance between two monomer considerably enhanced. Thus, in the composite system, the cargo follows the engine (JP) as if without facing any hydrodynamic interaction. 
  
\section{Conclusion} We numerically investigated details of escape kinetics of a Janus particle carrying a cargo  from the minimum of a meta-stable state. We exploit rigid dimer approximation (which allows us to reduce a two-particle problem to two one particle problems) to better understand the essential features of escape mechanisms. Three different regimes of the correlation time of self-propelled velocity  along with  hydrodynamic interaction have been analyzed separately. For very fast rotational diffusion, the escape rate can be simplified to a simple Kramers' problem of a passive particle with an effective temperature. On the other hand, when the rotational relaxation is very slow, escape rate of a rigid dimer can again be estimated based on conventional flux over population  method with an adiabatic approximation. In the intermediate regime of rotational relaxation,  a resonant activation like behavior, manifested by appearance of a minimum in $T_{MET}$ versus $\tau_r$. This type of resonance is attributed to synchronization between the noise-induced hopping event of the dimer and the orientational relaxation of the self-propelled velocity. 
     
We also extend our study to analyze escape kinetics when  segments of the dimer weakly interact with substrate. This situation corresponds to escape problem with low barrier in comparison to the self-propulsion strength. Very prominent resonance minima in $T_{MET}$ versus $\tau_r$ and the  linearly growth of mean exit time with $\tau_r$ in the asymptotic regime, are two apparent distinct features in this regime. Interestingly, the ratio of escape rates, $T_r$, passes through a maximum which falls far away from the synchronization points (either of dimer or the single JP) of resonant activation. Further, hydrodynamic interaction enhances escape rate to a considerable extent. Its effect is more pronounced at low self-propulsion strength and at some intermediate regime of rotational relaxation time. Important simulation results have been summarized in the Table 1.  We hope all these simulation results will be useful to synthesis suitable JPs with desire transport features and to operate them in a controlled manner for potential applications in targeted drug delivery, non-invasive surgery, cleaning biological channels, separating particles   etc., to name a few.

\section*{Acknowledgements}
T. D. thanks UGC, New Delhi, India, for the award of a Senior Research Fellowship.  P.K.G. is supported by SERB Start-up Research Grant (Young Scientist) No. YSS/2014/000853 and the UGC-BSR Start-Up Grant No. F.3092/2015.

\end{document}